\pdfoutput=1

\documentclass[11pt,twoside,a4paper,cmspaper,final,collab]{cms-tdr}
\def\svnVersion{185181}\def\cmsCernNo{2012-342}\def\cmsCernDate{\today}\def\cmsMessage{Submitted to Physics Letters B}

\begin{document}\cmsNoteHeader{EXO-11-035}

\hyphenation{had-ron-i-za-tion}
\hyphenation{cal-or-i-me-ter}
\hyphenation{de-vices}

\RCS$Revision: 134303 $
\RCS$HeadURL: svn+ssh://svn.cern.ch/reps/tdr2/notes/EXO-11-035/trunk/EXO-11-035.tex $
\RCS$Id: EXO-11-035.tex 134303 2012-07-02 14:01:11Z sigamani $
\newlength\cmsFigWidth
\ifthenelse{\boolean{cms@external}}{\providecommand{\cmsLeft}{top\xspace}}{\providecommand{\cmsLeft}{left\xspace}}
\ifthenelse{\boolean{cms@external}}{\providecommand{\cmsRight}{bottom\xspace}}{\providecommand{\cmsRight}{right\xspace}}
\ifthenelse{\boolean{cms@external}}{\setlength\cmsFigWidth{0.47\textwidth}}{\setlength\cmsFigWidth{0.49\textwidth}}
\ifthenelse{\boolean{cms@external}}{\renewcommand{\ETslash}{\ensuremath{E_{\mathrm{T}}\hspace{-0.95em}/\kern0.5em}\xspace}}{}

\cmsNoteHeader{EXO-11-035} 
\title{Search for long-lived particles in events with photons and missing energy in proton-proton collisions at $\sqrt{s} = 7\TeV$}

\date{\today}

\newcommand{\lumin}{\ensuremath{4.9 \pm 0.1\fbinv}\xspace}
\newcommand{\luminnoErr}{4.9\fbinv\xspace}
\newcommand{\NONE}{\PSGczDo\xspace}

\abstract{
Results are presented from a search for long-lived neutralinos
decaying into a photon and an invisible particle, a signature
associated with gauge-mediated supersymmetry breaking in supersymmetric
models. The analysis is based on a \luminnoErr\ sample of proton-proton
collisions at $\sqrt{s} = 7$\TeV, collected with the CMS detector at the
LHC. The missing transverse energy and the time of arrival of the photon
at the  electromagnetic calorimeter are used to search for an excess of
events over the expected background. No significant excess is observed,
and lower limits at the 95\% confidence level are obtained on the mass of
the lightest neutralino, $m>220$\GeV (for $c\tau<500$\unit{mm}), as well
as on the proper decay length of the lightest neutralino, $c\tau > 6000$\unit{mm}
(for $m< 150$\GeV).
}

\hypersetup{%
pdfauthor={CMS Collaboration},%
pdftitle={Search for long-lived particles in events with photons and missing energy in proton-proton collisions at sqrt(s) = 7 TeV},%
pdfsubject={CMS},%
pdfkeywords={CMS, LHC, SUSY, Exotica, GMSB, Long-Lived, Photons, Displaced}}

\maketitle 

\section{Introduction}

\begin{figure*}[tbh]
\begin{center}
\includegraphics[width=0.45\textwidth]{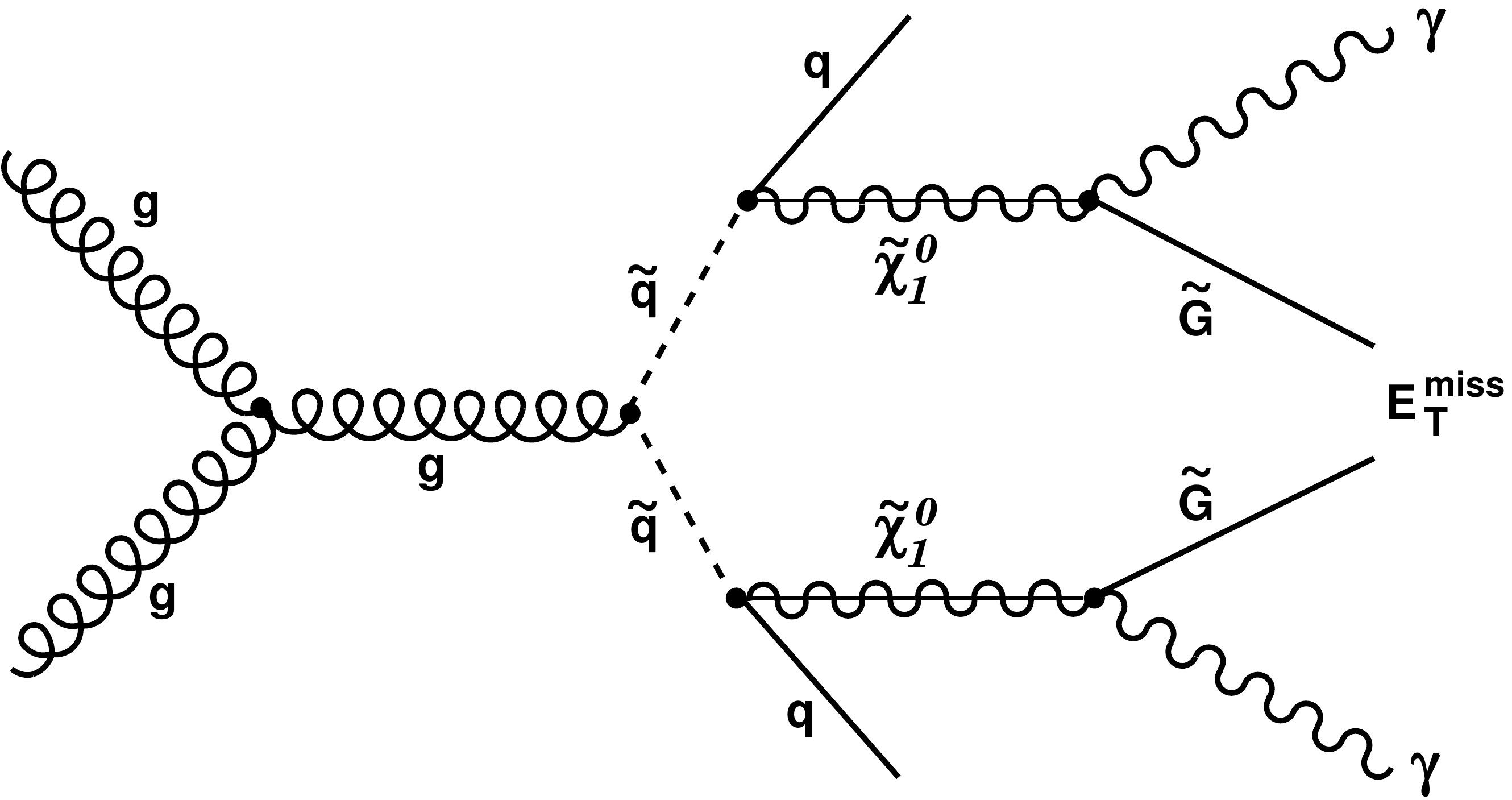}\quad
\includegraphics[width=0.45\textwidth]{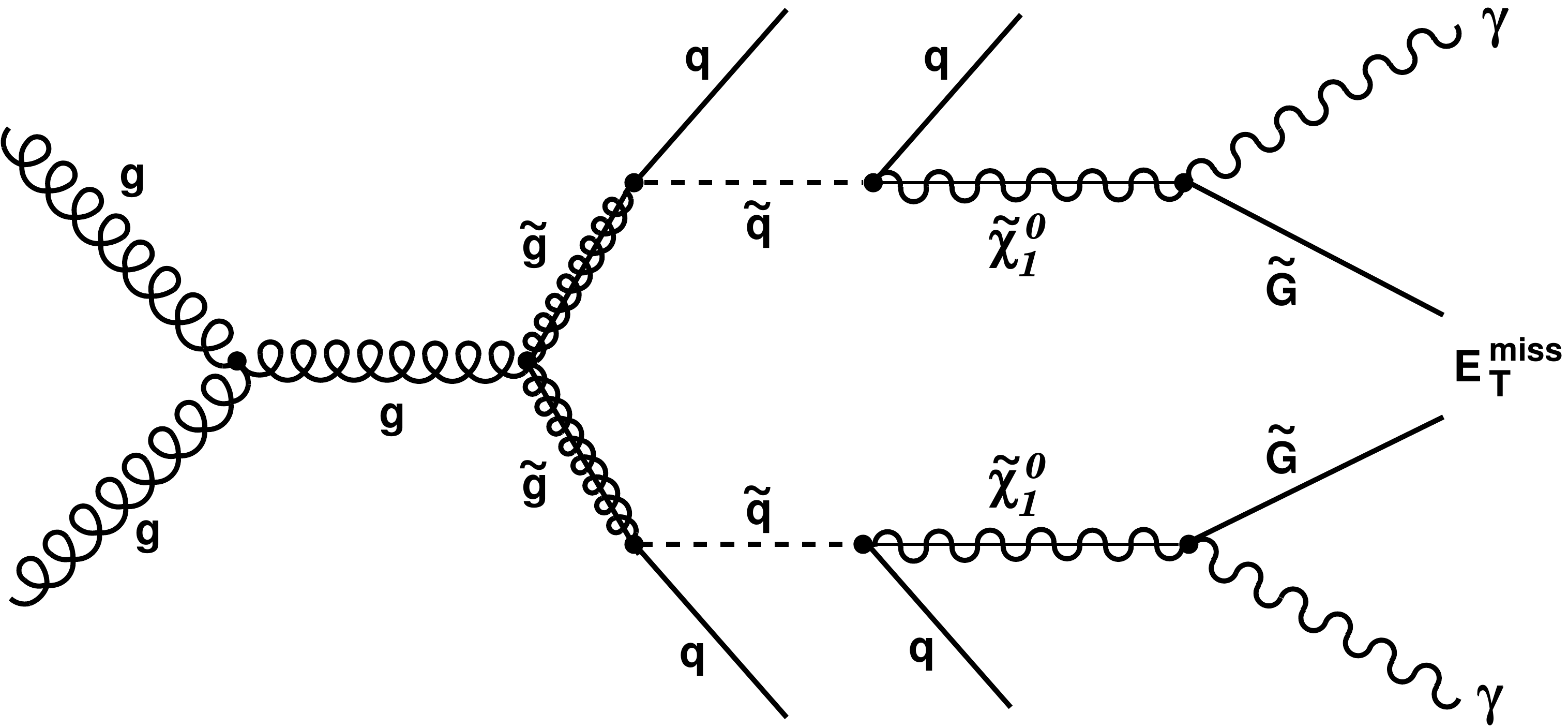}
\\ \vspace*{0.4cm}
\includegraphics[width=0.45\textwidth]{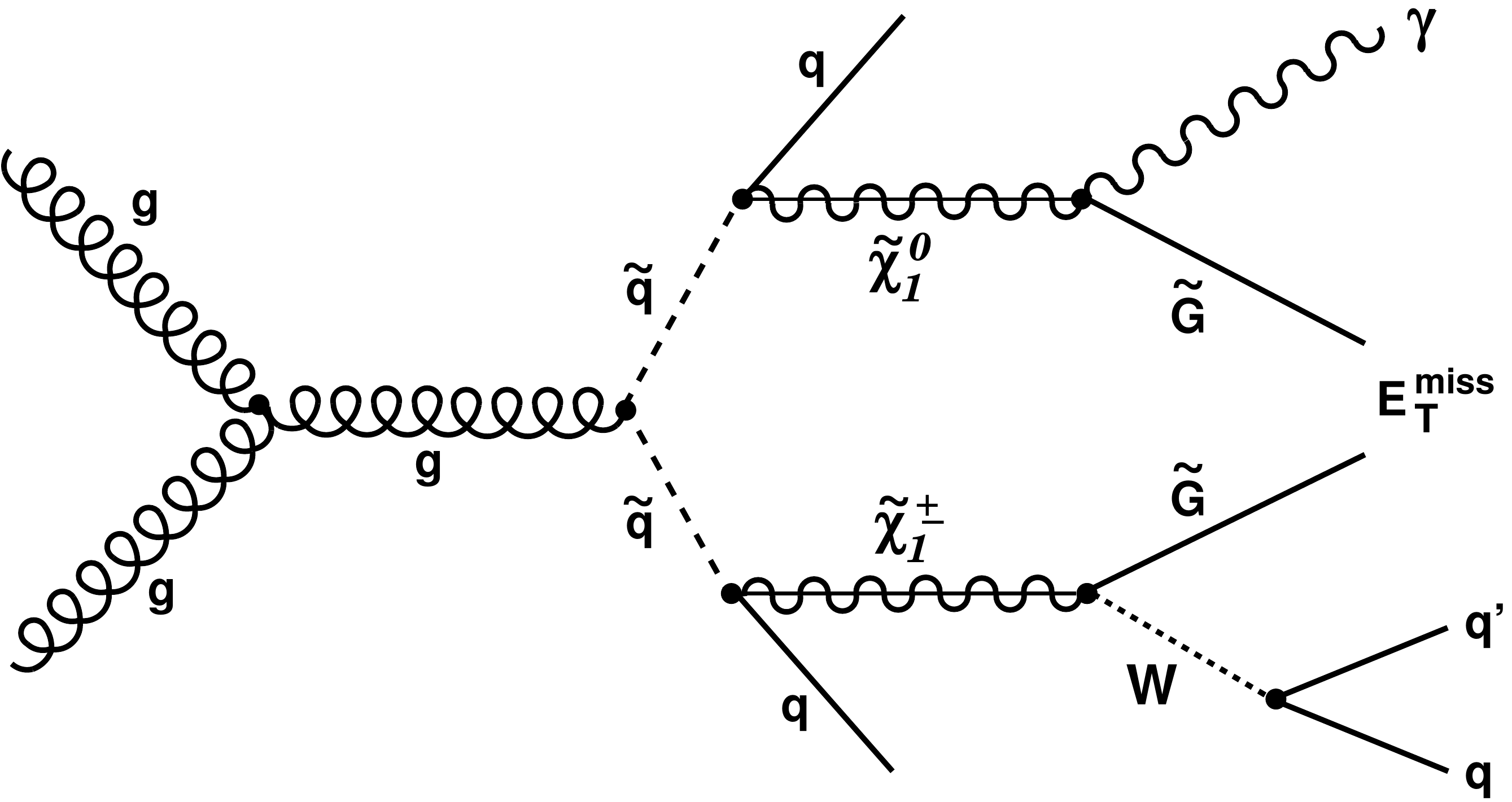}\quad
\includegraphics[width=0.45\textwidth]{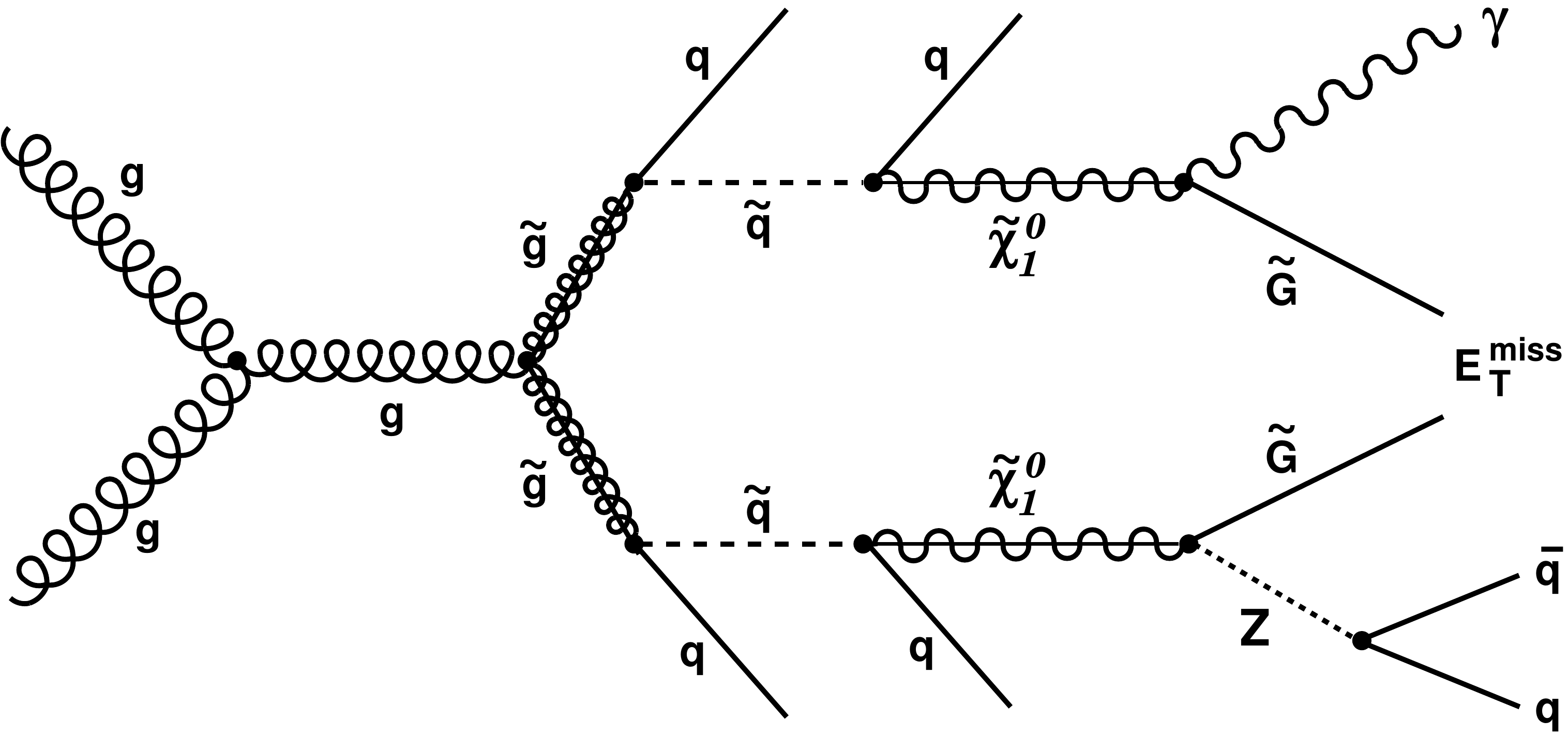}
\caption{\label{fig:diagrams} Example diagrams for SUSY
processes that result in a diphoton (top) and single-photon (bottom)
final state through squark (left) and gluino (right) production at the LHC.}
\end{center}
\end{figure*}

New, heavy particles with long lifetimes are predicted in many models of physics beyond the standard model (SM),
such as hidden valley scenarios~\cite{Strassler:2006im} or supersymmetry (SUSY) with gauge-mediated
supersymmetry breaking (GMSB)~\cite{Giudice:1998bp}.
Under the assumption of R-parity conservation~\cite{Farrar:1978xj},
strongly-interacting supersymmetric particles would be pair-produced at the
Large Hadron Collider (LHC).
The decay chain may include one or more quarks and gluons, as well as
the lightest supersymmetric particle (LSP), which escapes detection,
giving rise to a momentum imbalance in the transverse plane.
A GMSB benchmark scenario,
commonly described as `Snowmass Points and Slopes 8'
(SPS8)~\cite{Allanach:2002nj} is used as the reference in this search.
In this scenario, the lightest neutralino (\PSGczDo)  is the next-to-lightest supersymmetric
particle, and can be long-lived.
It decays to a photon (or a Z boson) and a gravitino (\PXXSG), which is
the LSP~\cite{Dimopoulos:1996vz}.
If \PSGczDo consists predominantly of the bino,  the superpartner of the $U(1)$ gauge field,
its branching fraction to a photon and gravitino is expected to be
large.
If \PSGczDo is wino-like, the superpartner of the $SU(2)$ gauge fields, its branching
fraction to a photon and gravitino is reduced.
Figure~\ref{fig:diagrams} shows several diagrams of possible squark and gluino pair-production processes
that result in a single-photon or diphoton final state.

The search criteria require only one identified photon in order to be sensitive to scenarios
with a large branching fraction for the neutralino decay to a Z boson and a gravitino.
For a long-lived neutralino, the photon from the $\PSGczDo\rightarrow\gamma\PXXSG$ decay
is produced at the \PSGczDo decay vertex, at some distance from the beam line,
and reaches the detector at a later time than the prompt, relativistic particles produced at the
interaction point.
In addition, the geometric shape of the energy deposit produced by such
photons is typically different from that of a prompt photon.
The time of arrival of the photon at the detector and the missing transverse energy are used
to discriminate signal from  background.

A search for a long-lived neutralino, decaying to a photon and a gravitino, is performed with
a novel technique using the excellent time measurement with the electromagnetic calorimeter.
Previous searches for long-lived neutralinos
have been performed by the CMS Collaboration~\cite{Chatrchyan:2012ir},
using the impact parameter of converted photons relative to the beam collision point,
and by the CDF~\cite{Aaltonen:2009tp} collaboration, using only the missing transverse energy in the event.
Other searches with  prompt photons, by the ATLAS~\cite{Aad:2012zj} and
D0~\cite{Abazov:2010us} collaborations,
place lower limits on the mass of the \PSGczDo at $280$\GeV and $175$\GeV, respectively,
in the SPS8 scenario, assuming $\mathcal{B}(\PSGczDo\rightarrow\gamma\PXXSG)=100\%$.

\section{Detector and data samples}

A detailed description of the Compact Muon Solenoid (CMS) detector can be found
elsewhere~\cite{Chatrchyan:2008aa}.
The detector's central feature is a superconducting solenoid providing a 3.8\unit{T} axial magnetic
field along the beam direction.
Charged particle trajectories are measured by a silicon pixel and strip tracker system
with full azimuthal coverage within  $|\eta| < 2.5$;
the pseudo-rapidity $\eta$ is defined as
$\eta = -\ln[\tan{(\theta/2)}]$, with $\theta$ being the polar angle with respect to the
counterclockwise beam direction.
A lead-tungstate (PbWO4) crystal electromagnetic calorimeter (ECAL)
and a brass/scintillator hadron calorimeter (HCAL) surround the tracker volume.
The ECAL is a high-granularity device. 
The barrel region consists of 61\,200 crystals with a frontal area of approximately
$2.2\times 2.2\unit{cm}^{2}$ corresponding to roughly  $0.0174\times0.0174$
in $\eta$-$\phi$ space.
Each of the two endcap sections consist of 3662 crystals with a frontal area of
$2.68 \times 2.68\unit{cm}^{2}$.
A typical shower spans approximately 10 crystals with energy deposits above the threshold.
Muons are measured in gas-ionization detectors embedded
in the steel return yoke of the magnet. The detector is nearly hermetic, allowing reliable
measurement of transverse momentum imbalance to be performed.
The time of arrival of electromagnetic particles
can be measured to excellent precision using the CMS ECAL~\cite{Chatrchyan:2009aj}.
The time reconstruction method is described in more detail in Section~\ref{subsec:obj}.

The analysis is performed on the proton-proton collision data
at a center-of-mass-energy of 7\TeV
recorded by the CMS detector at the LHC,
corresponding to an integrated luminosity of \lumin.
Events with at least one high transverse momentum ($\pt$) isolated photon
in the barrel region ($|\eta|<1.44$) and at least three jets in the final state are selected in this analysis.
The data were recorded using the CMS two-level trigger system.
Several trigger selections have been used
due to the increasing instantaneous luminosity during the data taking.
The first 0.20\fbinv of data were collected with a trigger requiring
at least one isolated photon with $\pt > 75$\GeV.
For the second 3.8\fbinv, the $\pt$ threshold was increased to $90$\GeV.
In the remaining 0.89\fbinv,
the trigger selection required at least one isolated photon
with $\pt > 90$\GeV in the  barrel region and
at least three jets with $\pt$ greater than 25\GeV.
All offline selection requirements are chosen to be more restrictive than the trigger selection.

Signal and background events are generated using Monte
Carlo (MC) packages \PYTHIA 6.4.22 \cite{Sjostrand:2006za} or \MADGRAPH 5~\cite{Alwall:2011uj}
with the CTEQ6L1~\cite{Pumplin:2002vw} parton distribution functions (PDFs).
The response of the CMS detector is simulated using the \GEANTfour package~\cite{Agostinelli:2002hh}.
Decays of secondary $\tau$ leptons, coming from  W and Z productions, are simulated with \TAUOLA~\cite{Was:2000st}.
The SUSY GMSB signal production follows the SPS8 proposal,
where the free parameters are the SUSY breaking scale ($\Lambda$) and the average
 proper decay length (c$\tau$) of the neutralino.
The $\PSGczDo$ mass explored is in the range of 140 to 260\GeV
(corresponding to $\Lambda$ values from 100 to 180\TeV),
with proper decay lengths ranging from c$\tau$ = 1\unit{mm} to 6000\unit{mm}.
These free parameters are varied to cover the range of experimental phase space allowed
by inner radius of the barrel section of the ECAL (1.29\unit{m}).

There is a non-negligible probability that several collisions may occur in a single bunch crossing
due to the high instantaneous luminosities at the LHC.
The presence of multiple interaction vertices in an event (pile-up)
affects the resolution of the transverse momentum measurement and the
performance of photon isolation requirements.
To account for the effects of pile-up, simulated events are re-weighted
so that the distribution of the number of interaction vertices
matches that in the data.

\section{Analysis technique}
This section, outlining the analysis technique,
starts with a description of the physics object reconstruction
followed by a brief explanation of the event selection criteria.
Finally, the definitions
of the key discriminating variables related to the ECAL cluster shape and the time of impact of the photon
on the surface of the ECAL are discussed.
The signal and background yields are determined with a binned maximum likelihood fit to
the two-dimensional distribution in these variables.

\subsection{Object reconstruction}
\label{subsec:obj}

Photons are reconstructed by identifying energy deposits in the
ECAL using the method explained in Ref.~\cite{CMS-PAS-EGM-10-006}.
Photons that are found to have converted into an electron-positron pair
in the detector material are not used in the analysis.
Electron or positron candidates are reconstructed starting from a cluster of energy deposits in the ECAL which is
then matched to the momentum associated with a track in the silicon tracker.
Electron candidates are required to have $|\eta|  < 1.44 $ or $1.56 < |\eta|  < 2.5$ to avoid
the region of transition between the barrel and endcap sections.
Photons are required to be spatially separated from electrons by at least
$\Delta R = \sqrt{(\Delta\eta)^{2}+(\Delta\phi)^{2}} = 0.25$,
where $\Delta\eta$ and $\Delta\phi$ are, respectively, differences between the photon
and the electron directions in pseudorapidity and azimuthal angle.

Jets are reconstructed from objects identified using the Particle-Flow (PF) algorithm~\cite{PFT-10-002}
with anti-\kt clustering~\cite{Cacciari:2008gp} and a distance parameter of 0.5.
In this analysis, the missing transverse energy (\ETslash) is defined as the
magnitude of the vector sum of the transverse momentum of all particles
identified in the PF algorithm in the event excluding muons.

\begin{figure}[tbhp]
  \begin{center}
    \includegraphics[width=\cmsFigWidth]{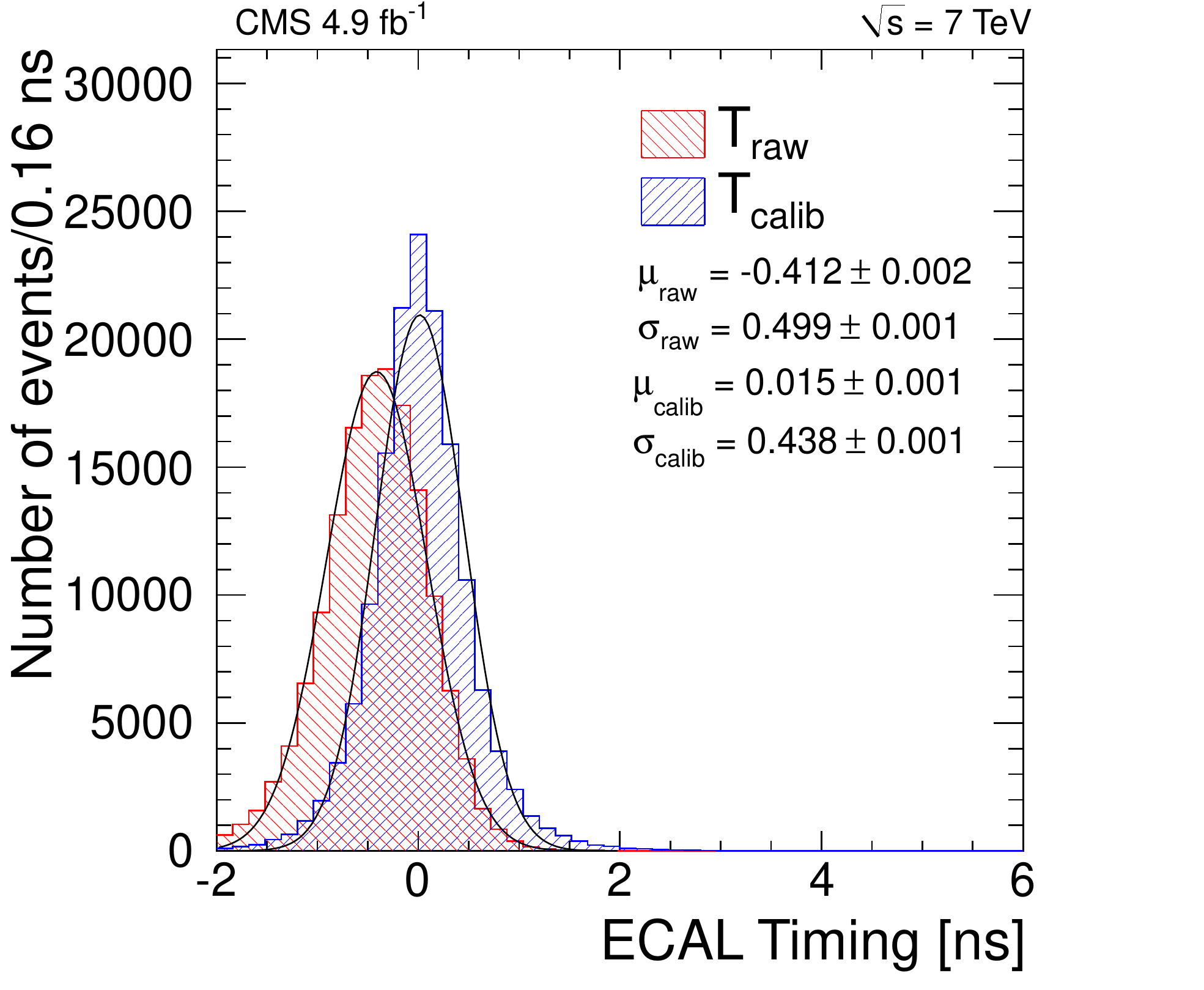}
        \caption{The ECAL timing distribution for data, before and after calibration,
                 overlaid with the results of the Gaussian fits.}
    \label{fig:timing1}
  \end{center}
\end{figure}

The time of impact, $T_\text{raw}$, for the photon on the surface of the ECAL
is the weighted time of impact measured in the crystals
within the cluster associated with  a photon candidate.
An event-by-event correction ($T_\text{prompt}$) is applied to $T_\text{raw}$ to account for
possible biases  due to the jitter in the trigger system, and to the imperfect knowledge of
the time of the interaction within the  bunch crossing.
This correction is computed using the time of impact of all crystals in the event,
excluding those belonging to the two most energetic photon candidates,
which are typically due to  prompt jets, low-energy prompt photons,
and photons from $\pi^{0}$ and $\eta$ decays.
The new calibrated ECAL timing is defined as
$T_\text{calib} = T_\text{raw} - T_\text{prompt}$.
With this definition, a particle produced at the interaction point
has a time of arrival of zero, whereas a delayed photon has a non-zero $T_\text{calib}$.
The distributions in data for $T_\text{raw}$ and $T_\text{calib}$, after the nominal selection,
are shown in Fig.~\ref{fig:timing1}.
The width of the main, Gaussian, component of $T_\text{calib}$ is slightly
smaller than that of $T_\text{raw}$, while there is some increase in the tails.
For the dominant background processes, the tails are taken into account by using control samples in data,
as described in Section~\ref{sec:bkg}.
In the determination of the yield, the distribution of $T_\text{calib}$ in
simulated signal events is used as a template for the signal contribution.
This distribution is narrower in simulation than in the data, because the
uncertainties in the time inter-calibration constants are not emulated.
A convolution with a Gaussian, whose parameters vary as a function of the photon energy, is performed
to reproduce the $T_\text{calib}$ resolution observed in data.

\begin{figure}[tbh]
  \begin{center}
\includegraphics[width=\cmsFigWidth]{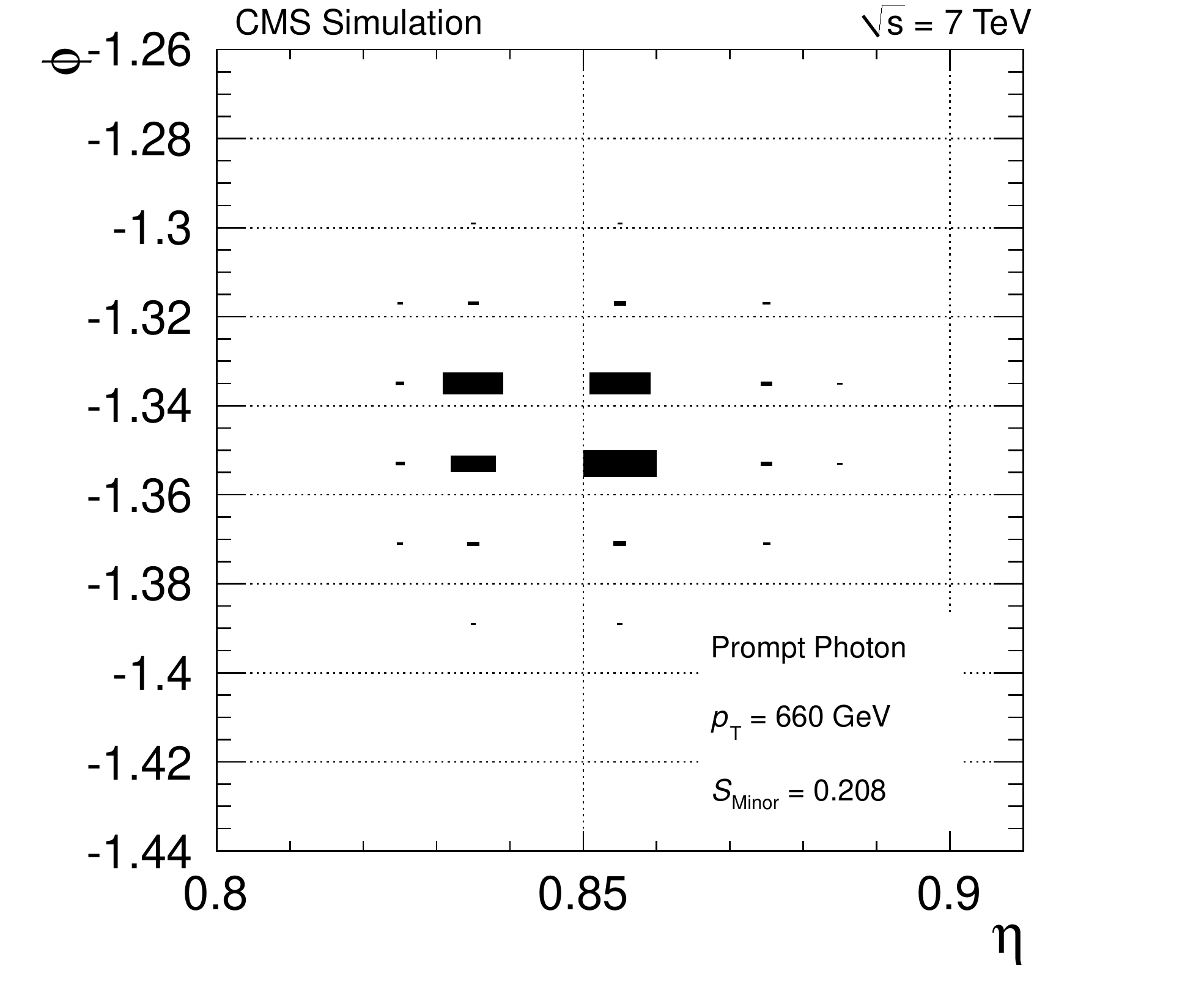}
\includegraphics[width=\cmsFigWidth]{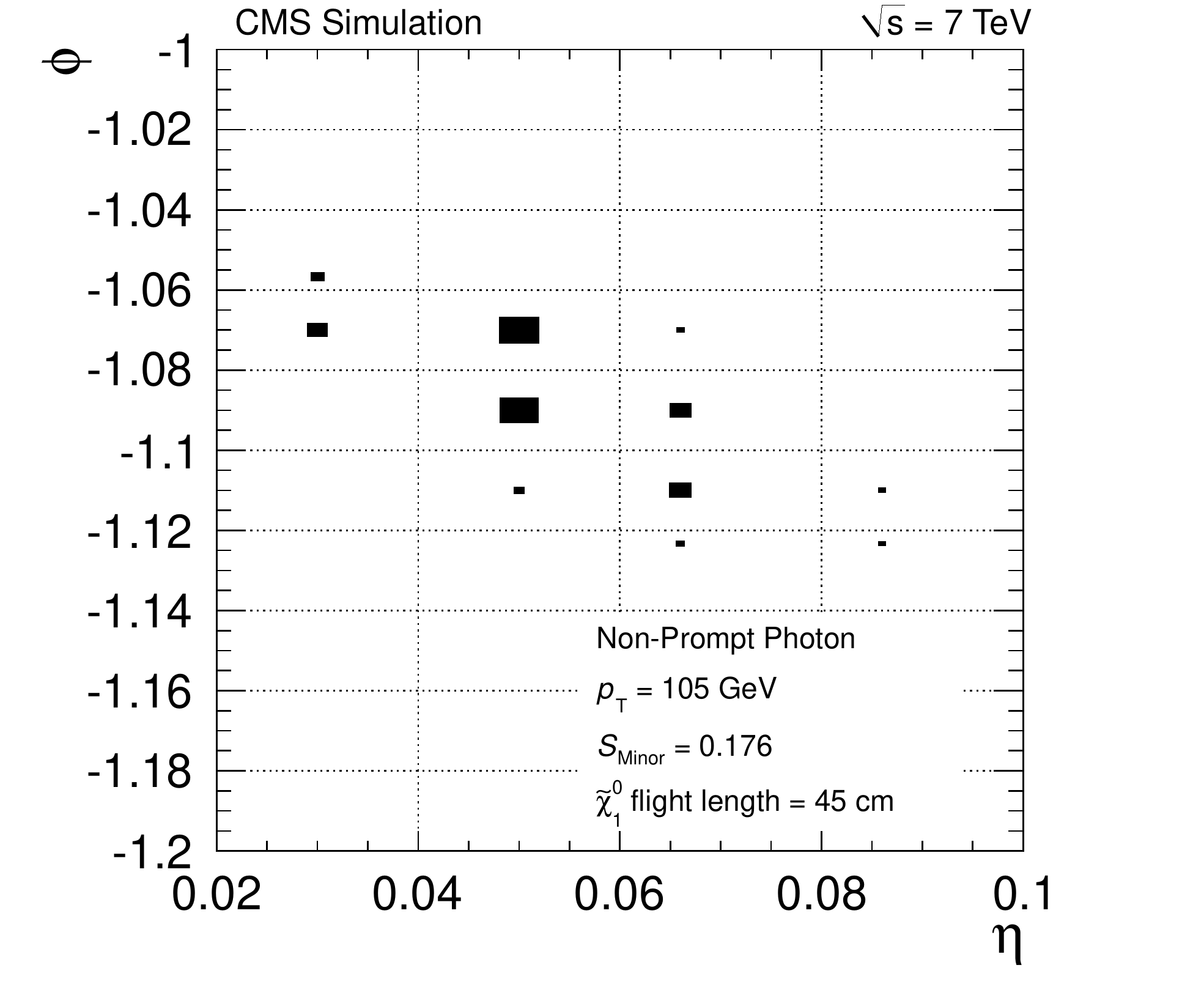}
        \caption{The distribution of energy deposition in the ECAL crystals for
                 a prompt (\cmsLeft) and a non-prompt (\cmsRight) photon.
                 Each rectangle represents an ECAL crystal and has a size that is proportional to the
                 energy deposited in that crystal. The non-prompt illustration is for a  \PSGczDo
                 flight length of $45$\unit{cm}.
                 }
    \label{fig:n-alpha}
  \end{center}
\end{figure}

\begin{figure}[tbh]
  \begin{center}
\includegraphics[width=\cmsFigWidth]{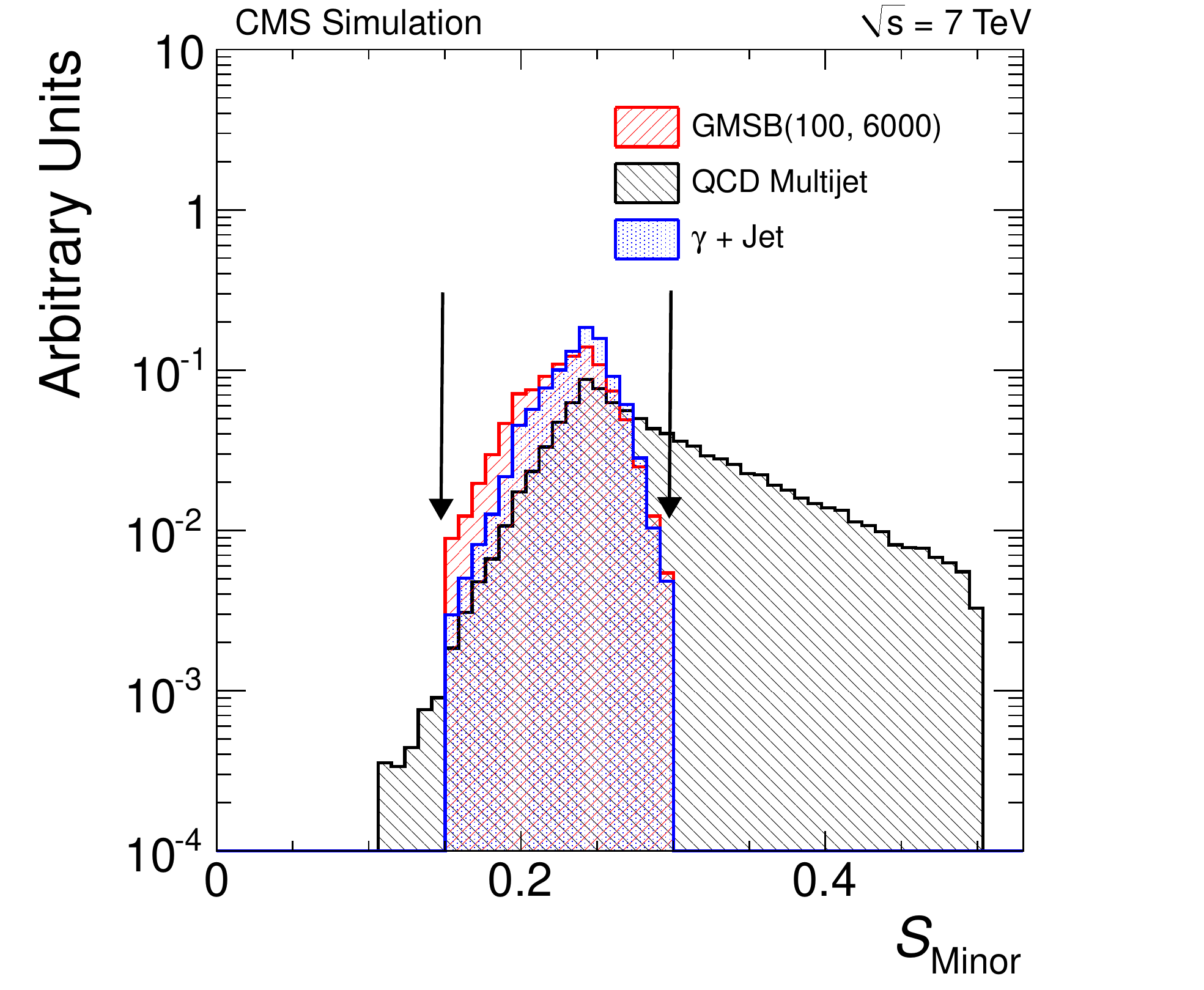}
        \caption{Normalized distribution of $S_\text{Minor}$ for simulated signal, $\gamma + \text{jets}$, and QCD multijet events. 
                 The arrows indicate the $S_\text{Minor}$ selection interval.}
    \label{fig:smin}
  \end{center}
\end{figure}

One of the distinctive features of a photon is the shape of the energy deposits it leaves in the ECAL.
Prompt photons have a roughly circular projected energy deposit on the ECAL surface,
while the energy deposits from jets typically have a larger width along the $\eta$ direction.
Non-prompt photons are expected to have an elliptical shape along an arbitrary direction,
as illustrated in Fig.~\ref{fig:n-alpha}, therefore the width of the energy deposit along
the $\eta$ direction is not optimal for the discrimination of jets.
In this search, the shape of the energy deposit is characterized by the minor axis ($S_\text{Minor}$) of its
projection on the internal ECAL surface.
The axis $S_\text{Minor}$ is computed using the geometrical properties of the distribution of the energy
deposit, and is defined as
\begin{equation}
S_\text{Minor} = \frac{ \left( S_{\phi\phi} + S_{\eta\eta} \right) - \sqrt{ \left( S_{\phi\phi} - S_{\eta\eta}  \right)^{2} + 4S^{2}_{\phi\eta}  }} {2},
\label{eq5.11}
\end{equation}
where $ S_{\phi\phi}$, $S_{\eta\eta}$, and $S_{\phi\eta}$ are the second moments of the
spatial distribution of the energy deposit in the ECAL in $\eta$-$\phi$ coordinates.
A large fraction of QCD multijet events can be rejected by applying requirements on $S_\text{Minor}$
as illustrated in Fig.~\ref{fig:smin}, where the normalized distributions of $S_\text{Minor}$ for
simulated signal and QCD multijet background events are shown.

\subsection{Event selection}

Events must have a primary vertex with at least four associated tracks and a position less than $2$\unit{cm}
from the center of the CMS detector in the direction transverse to the beam and $24$\unit{cm} in
 the direction along the beam.
Events are also required to have at least three jets with $\pt > 35$\GeV
and spatially separated from photons by at least $\Delta R = 0.5$.

Photon candidates are required to have $\pt \ge 100$\GeV and $|\eta|\le 1.44$ and
to be isolated in the HCAL, the ECAL, and the tracker.
An absolute  isolation parameter is defined as the scalar sum of
the transverse energies of tracks or calorimeter deposits
in a cone of aperture 0.3 around the photon direction,
excluding the contribution from the photon itself.
A relative isolation parameter is defined as the ratio of the absolute isolation and the photon $\pt$.
In the tracker, the relative isolation is required to be
less than 0.1. In the ECAL and the HCAL, the relative isolation is required to be less than 0.05
and the absolute isolation less than 2.4\GeV.
Thresholds on both absolute and relative isolation are set in the ECAL and HCAL to avoid imposing
requirements that are more restrictive than the noise level.
The energy deposit by a photon candidate is required to  have $0.15 < S_\text{Minor} <0.30$. This requirement
is optimized to select candidates that are more likely to be real photons.

The signal efficiencies for selecting one photon and at least three jets
are summarized in Table \ref{tab:PhotonEfficiency1} for proper decay lengths between $1$\unit{mm} and $6000$\unit{mm}
and for $\Lambda$ between $100$\TeV and $180$\TeV. The efficiency drops by a factor of two
between $c \tau= 1$\unit{mm} and $6000$\unit{mm},  since, with increasing decay time, the probability
of the \PSGczDo to decay outside the detector is enhanced.
\begin{table*}[tbh]
\centering
      \topcaption{Selection efficiency in percent.
            The reported uncertainties include the contributions of
             systematic effects, for various signal samples.
     }
\begin{tabular}{c | c |c c c c}
\hline
{   $\Lambda$ (\TeVns{})} & {    $M_{\PSGczDo}$ (\GeVns{})} & {   $c \tau  =1$\unit{mm}} & {    $c \tau  =250$\unit{mm}} & {    $c \tau  =2000$\unit{mm}} & {    $c \tau  =6000$\unit{mm}}  \\
\hline
{  100} & {  140} & 18.7 $\pm$ 0.3        & 18.4 $\pm$ 0.2       &8.4  $\pm$ 0.2 &3.3  $\pm$ 0.1\\
{  120} & {  170} & 24.9 $\pm$ 0.4        & 24.6 $\pm$ 0.3       &15.1 $\pm$ 0.4 &6.6  $\pm$ 0.1\\
{  140} & {  200} & 30.4 $\pm$ 0.3        & 31.3 $\pm$ 0.3       &22.2 $\pm$ 0.4 &11.4 $\pm$ 0.3\\
{  160} & {  230} & 35.5 $\pm$ 0.3        & 36.1 $\pm$ 0.6       &29.4 $\pm$ 0.4 &17.0 $\pm$ 0.4\\
{  180} & {  260} & 40.1 $\pm$ 0.7        & 38.0 $\pm$ 0.5       &36.0 $\pm$ 0.5 &22.2 $\pm$ 0.4\\
\hline
\end{tabular}
\label{tab:PhotonEfficiency1}
\end{table*}

\section{Background estimation}
\label{sec:bkg}

The primary sources of background in the analysis are QCD multijet
events and $\gamma + \text{jets}$ events, which together make
up 99\% of the sample.
Improper reconstruction of jets can give rise to fake missing
transverse energy, while photons produced in the decays of hadrons
(mostly energetic $\Pgpz$ and $\eta$) can sometimes  pass the isolation
criteria.

A large fraction of $\gamma + \text{jets}$ events, characterized by a smaller jet multiplicity
compared to signal, are rejected by requiring at least three jets in the event.
The residual contribution of these backgrounds is estimated from the data.

In addition, there are other (non-QCD) processes with genuine $\ETslash$, largely  comprised of
$\PW/\cPZ + \gamma + \text{jets}$ and \ttbar\ events, where the W boson decays into a lepton and a neutrino.
There is also a small contribution from Drell--Yan processes.
These events make up less than 1\%  of the total sample
but are taken into account since they can play a role in the
tails of the \ETslash\ distribution where signal is expected.
Simulated events are used to estimate the contribution of these processes.

Finally, additional backgrounds from events not originating from proton-proton collisions,
including cosmic rays and beam-halo muons, are also expected.
The contribution of these events is reduced to negligible levels by requiring
$T_\text{calib}$ of the most energetic photon candidate to be greater than $-2$\unit{ns}, and
the event to have an identified primary vertex and at least three jets.

Because of the difficulty of accurately predicting cross sections and jet multiplicities
for multijet and $\gamma +\text{jets}$ processes, their contribution is estimated with methods based on the data.
The QCD multijet control sample is obtained by selecting events with at least three jets
and a photon candidate passing a less stringent identification requirement but failing the
nominal photon selection criteria.
The $\gamma + \text{jets}$ control sample consists of events with one photon which satisfies the nominal selection.
Events with the angle in the transverse plane between the highest-$\pt$ jet (leading jet) and the photon
smaller than $2/3 \,\pi$ are rejected.
The ratio of the transverse momenta of the leading jet to
that of the photon is required to be between 0.6 and 1.4,
while for the subleading jet the ratio is required to be less than 0.2.
The contribution of non-QCD and signal events to these two control samples is estimated to be, respectively, 1\% and
less than 0.01\%.

\begin{figure}[ptbh]
  \begin{center}
    \includegraphics[width=\cmsFigWidth]{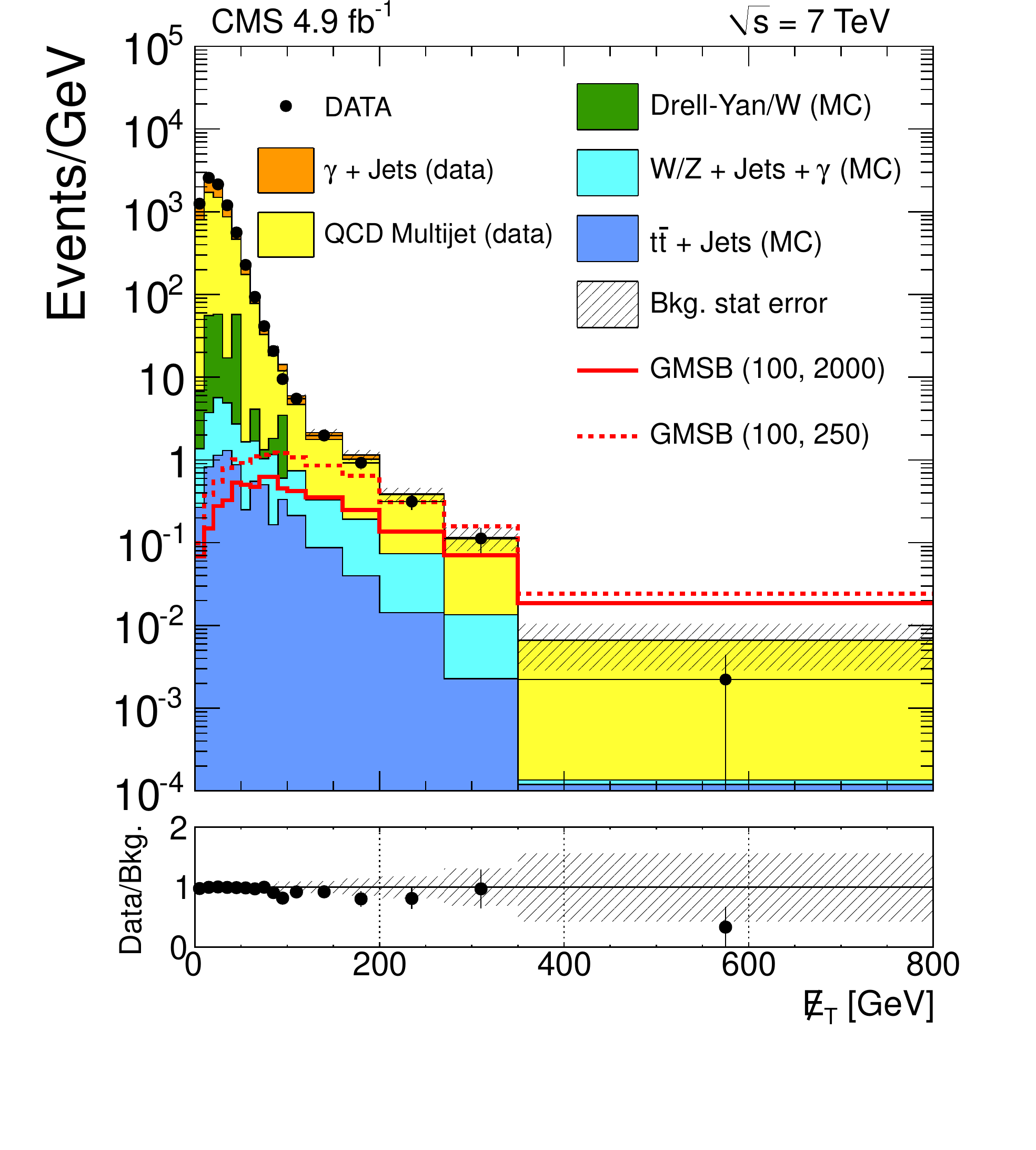}
    \includegraphics[width=\cmsFigWidth]{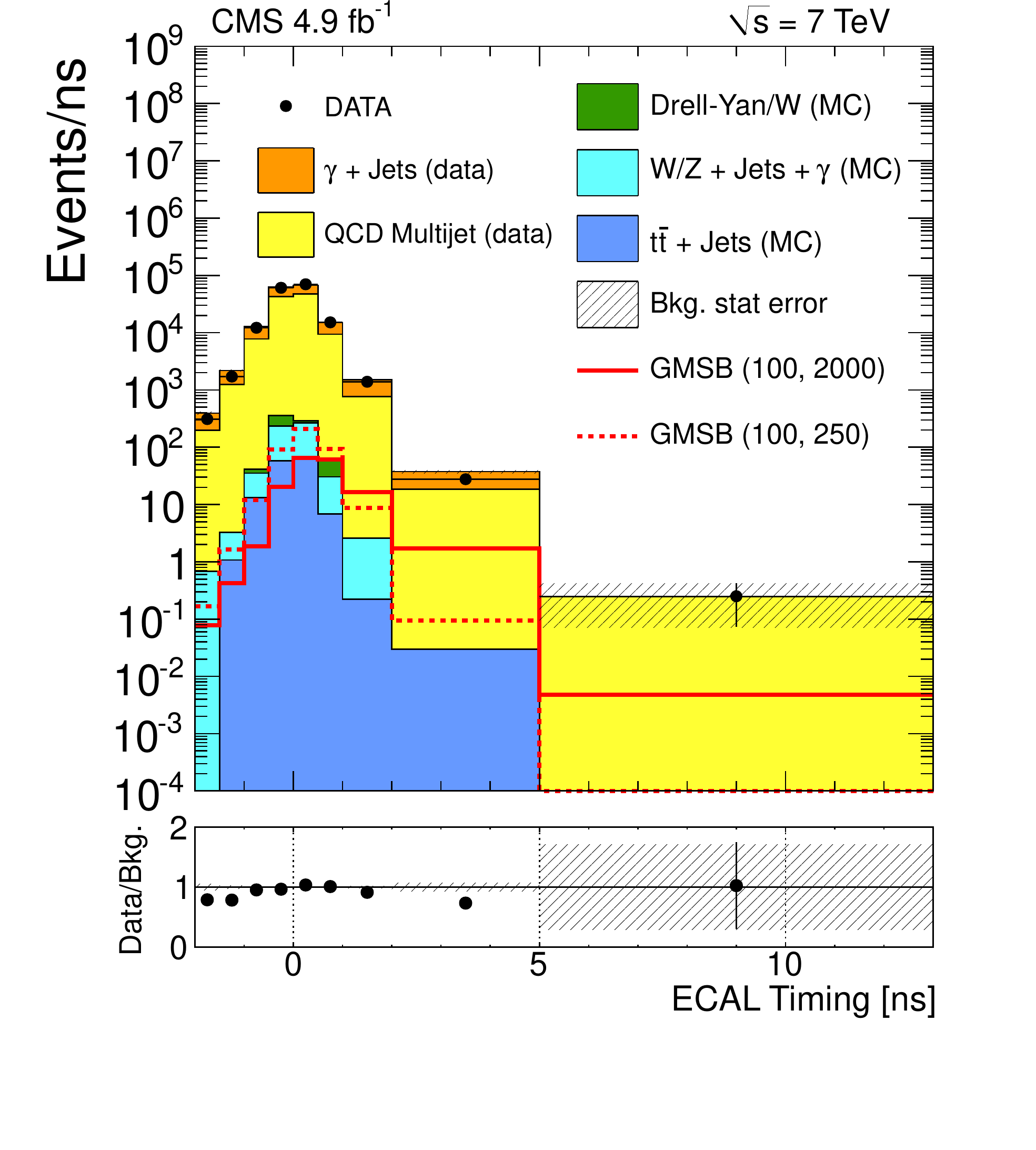}
    \caption{The one-dimensional projection for \ETslash\ (\cmsLeft) and for ECAL timing (\cmsRight),
         after all selection requirements. The multijet and $\gamma + \text{jets}$ backgrounds
         are normalized to the yields from the fit. The rest of the backgrounds
         are fixed according to the integrated luminosity of the data.
         The GMSB(100,2000) benchmark point corresponds to $\Lambda =100$\TeV, $c\tau=2000$\unit{mm} and
         the GMSB(100,250) benchmark point corresponds to $\Lambda =100$\TeV, $c\tau=250$\unit{mm}.}
    \label{fig:results2}
  \end{center}
\end{figure}

\begin{figure}[tbh]
  \begin{center}
    \includegraphics[width=\cmsFigWidth]{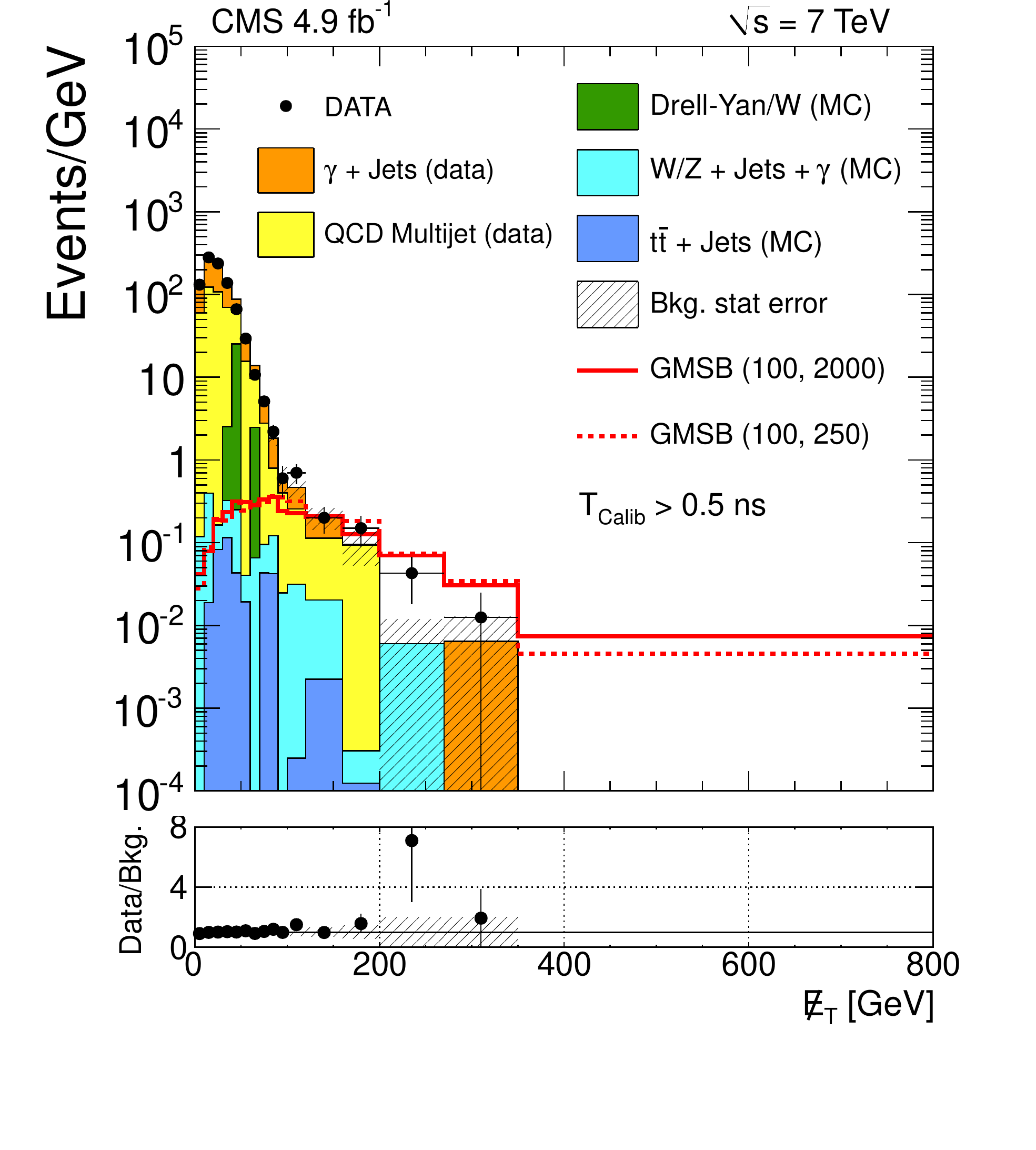}
    \includegraphics[width=\cmsFigWidth]{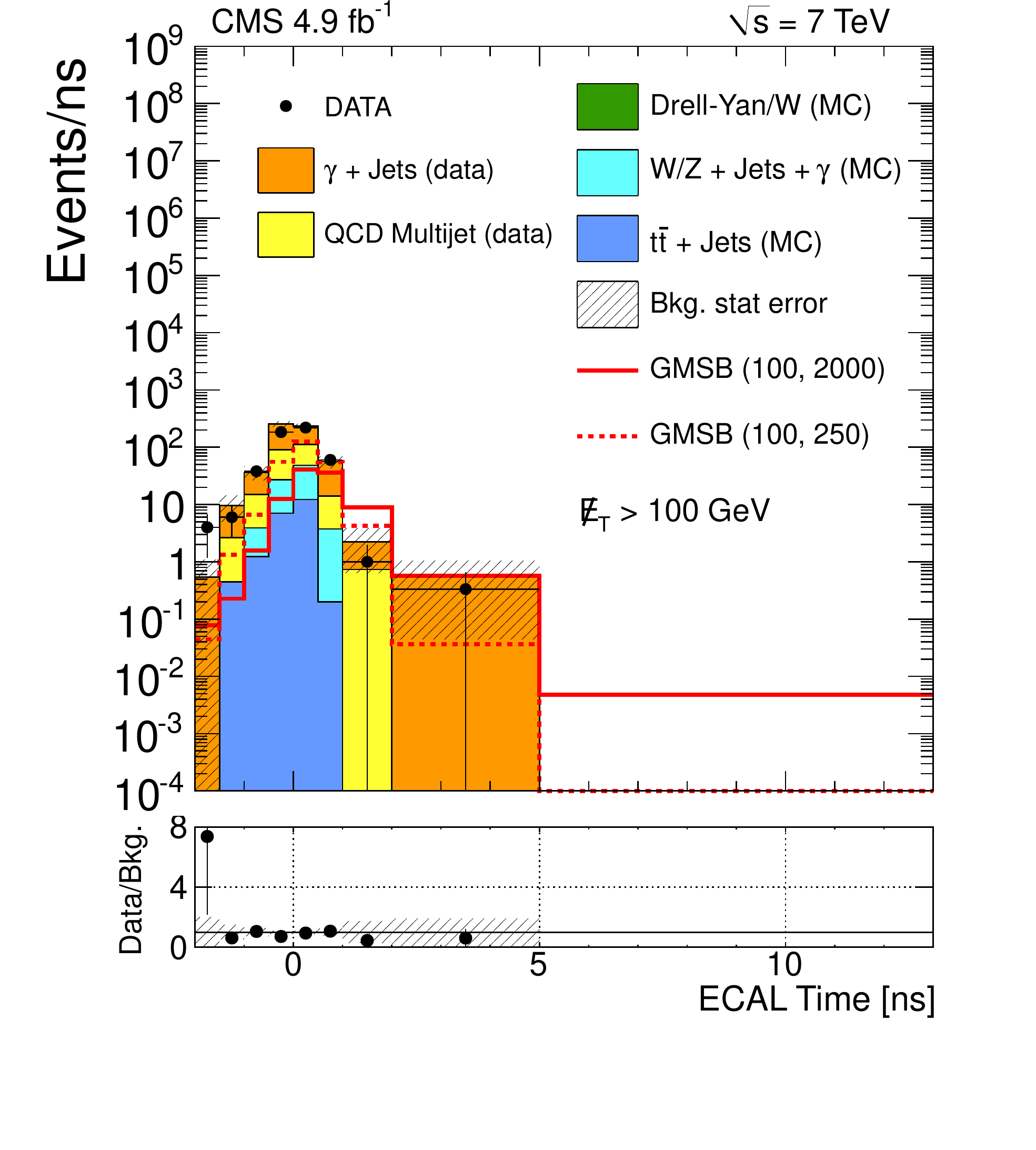}
    \caption{The one-dimensional projection after all selection requirements
         for \ETslash\ for events with $T_\text{calib} > 0.5\unit{ns}$ (\cmsLeft) and
         for ECAL timing (\cmsRight) for events with $\ETslash>100\GeV$.
         The multijet and $\gamma + \text{jets}$ backgrounds
         are normalized to the yields from the fit. The rest of the backgrounds
         are fixed according to the integrated luminosity of the data.
         The GMSB(100,2000) benchmark point corresponds to $\Lambda =100$\TeV, $c\tau=2000$\unit{mm} and
         the GMSB(100,250) benchmark point corresponds to $\Lambda =100$\TeV, $c\tau=250$\unit{mm}.}
    \label{fig:met-largetime}
  \end{center}
\end{figure}

To estimate the number of background and signal events in data,
a maximum likelihood fit is performed to the two-dimensional distribution of \ETslash\ and $T_\text{calib}$.
The correlation coefficient between \ETslash\ and $T_\text{calib}$ is $0.05$ for events with $\ETslash > 100\GeV$
and $T_\text{calib}>0.5\unit{ns}$, and $0.001$ when all events are considered.
Binned shape templates are derived from simulated events for signal and non-QCD backgrounds.
Templates for QCD multijet and $\gamma + \text{jets}$  are derived from data control samples as described earlier.
The relative normalization of the QCD multijet and $\gamma + \text{jets}$ components is fixed to 67\% and 33\%,
respectively, based on studies with simulated events.
The normalization of the non-QCD templates are fixed in the fit
according to the measured cross sections (statistical uncertainties in the cross sections are less than 3\%) and
the integrated luminosity of the data sample.
Studies have been performed with pseudo-experiments to confirm the stability of the fit
and to verify that the fit results are unbiased.
The measured signal and background yields in data, obtained with the likelihood
fit, are summarized in Table~\ref{tab:FinalNumbers}.
The one-dimensional projections of \ETslash\ and $T_\text{calib}$ for the data and expected
backgrounds, as determined from the fit, are illustrated in Fig.~\ref{fig:results2}. No excess of events
is observed beyond the SM backgrounds and the fitted signal yield is compatible with zero.
It should be noted that the discriminating power of individual variables is not apparent in these projections
because the largest sensitivity to signal is in the region with both large \ETslash\ and large $T_\text{calib}$.
The improved background discrimination is visible in Fig.~\ref{fig:met-largetime}
where  the one-dimensional projection of \ETslash\ for events with $T_\text{calib} > 0.5$\unit{ns} is illustrated.
\begin{table}[tbh]
\centering
 \topcaption{The measured signal and background yields determined with the maximum likelihood fit to the data.
The relative composition of QCD multijet and $\gamma + \text{jets}$ backgrounds have been normalized to 67\% and 33\% with
respect to each other.
The expected signal yields are $211$ events for the GMSB(100,250) benchmark point  and
$96$  for GMSB(100,2000).
The GMSB(100,250) benchmark point corresponds to $\Lambda =100$\TeV, $c\tau=250$\unit{mm} and
the GMSB(100,2000) benchmark point corresponds to $\Lambda =100$\TeV, $c\tau=2000$\unit{mm}.
The reported uncertainties are statistical only and are determined in the fit.}
\begin{tabular}{c c }
\hline
                                         		  & { Events}  \\
\hline
{  GMSB (100, 250)}                          & $ 6  \pm 8$ \\
{  GMSB (100, 2000)}                         & $ 4 \pm 4$\\
{  QCD multijet and $\gamma + \text{jets}$}               & $ 80\,900 \pm 300$\\
{  \ttbar\ + jets (fixed)}                 & $ 73$ \\
{  $\PW \to \Pe \nu + \text{jets}$ (fixed)}              & $ 116 $ \\
{  Drell--Yan + jets (fixed)}                  & $ 67 $ \\
{ $\PW/\cPZ + \text{jets} + \gamma$ (fixed)}              & $ 215 $ \\
\hline
{  Total background}			  & $81\,400 \pm 300$ \\
\hline
{ Data}                                     & $81\,382$ \\
\hline
\end{tabular}
\label{tab:FinalNumbers}
\end{table}

\section{Systematic uncertainties}

Several sources of systematic uncertainty have been considered and their contributions
are summarized in Table~\ref{tab:Uncertainty}.
The largest single contribution to the systematic uncertainties derives from the uncertainty in the modeling
of the  background shape.
A bin-by-bin  variation of the background shape template according to the Poisson uncertainty is used to determine
the contribution of each type of background.
An additional uncertainty is assessed for the QCD multijet and $\gamma + \text{jets}$ processes using simulated events,
by comparing the shapes of \ETslash\ and $T_\text{calib}$ for the control sample
and for a sample obtained with the nominal selection criteria.
The difference observed in simulation is used to re-weight the shapes obtained
in data control samples. The dominant contribution is due to the difference in the \ETslash\ distributions.
The small tails in the distribution of $T_\text{calib}$ are accounted for by using data control samples
to derive the templates, rather than relying on simulation.
The uncertainty in the relative fraction of QCD multijet and $\gamma + \text{jets}$ events is estimated to be 33\%.
The main contribution to this uncertainty is due to the next-to-leading correction for the $\gamma + \text{jets}$ cross section. 
Additional contributions are included to take into account the 
the observed difference between the number of events in the $\gamma + \text{jets}$ control sample in data
and the expected number of events according to $\PYTHIA$ (10\%), and to the QCD multijet events misidentified 
as $\gamma + \text{jets}$ events (10\%).

The main contributions to the uncertainty in the signal shape modeling derive from
the uncertainty in the \ETslash\ resolution and the determination of $T_\text{calib}$.
The contribution of the $\ETslash$ resolution uncertainty is estimated by smearing the \ETslash\
distribution of simulated signal events.
A systematic uncertainty of 0.1\unit{ns} is assigned to the measurement of the time of impact $T_\text{calib}$.
This uncertainty is determined using a sample of $\gamma + \text{jets}$ events by
measuring the difference between the average $T_\text{calib}$ values in data and simulation,
as a function  of the photon $\pt$.

The uncertainty in the luminosity determination is 2.2\%~\cite{CMS-PAS-SMP-12-008}.
The remaining sources of systematic uncertainty affecting the signal acceptance are the following.
The calorimeter response to different types of particles is not perfectly linear and hence
corrections are made to properly map the measured jet energy deposition.
The uncertainty on this correction is referred to as the uncertainty on the jet energy scale
and varies as a function of position and transverse momentum of the jet.
Similarly, the uncertainty on the photon energy scale in the barrel is estimated to be 1.0\%,
based on the final-state radiation measurement with Z bosons~\cite{Chatrchyan2011535}.
Following the recommendations of the PDF4LHC group~\cite{Botje:2011sn}, PDF and the strong
coupling constant ($\alpha_\mathrm{s}$) variations
of the MSTW2008~\cite{Martin:2009iq}, CTEQ6.6~\cite{Nadolsky:2008zw} and NNPDF2.0~\cite{Ball:2010de}
PDF sets are taken into account and their impact on the signal acceptance is estimated.
\begin{table}[tbh]        \centering
\topcaption{Summary of the systematic uncertainties in
the background and signal shapes, as well as in the signal acceptance $\times$ efficiency.
The signal uncertainties are evaluated individually for every signal point,
although only the maximum and minimum values associated with each source are  quoted.}
\begin{tabular}{c | c}
\hline
{ Source }                                  & { Uncertainty (\%)} \\
\hline
\multicolumn{2}{c}{Background} \\
\hline
Shape             			& 10  \\
Normalization     			& 0.3  \\
Multijet/$\gamma + \text{jets}$ fraction      	& 0.8  \\
\hline
\multicolumn{2}{c}{Signal shape} \\
\hline
\ETslash\ resolution                        & 0.2--2  \\
ECAL timing uncertainty                  & 1--5  \\
\hline
\multicolumn{2}{c}{ Signal acceptance $\times$ efficiency} \\
\hline
Photon energy scale                     & 0.5--3 \\
Jet energy scale                        & 0.02--0.05  \\
Jet energy resolution                   & 0.01--2  \\
PDF uncertainties                       & 0.1--2  \\
\hline
\end{tabular}

     \label{tab:Uncertainty}
\end{table}

\section{Results}

\begin{figure}[tbhp]
  \begin{center}
    \includegraphics[width=\cmsFigWidth]{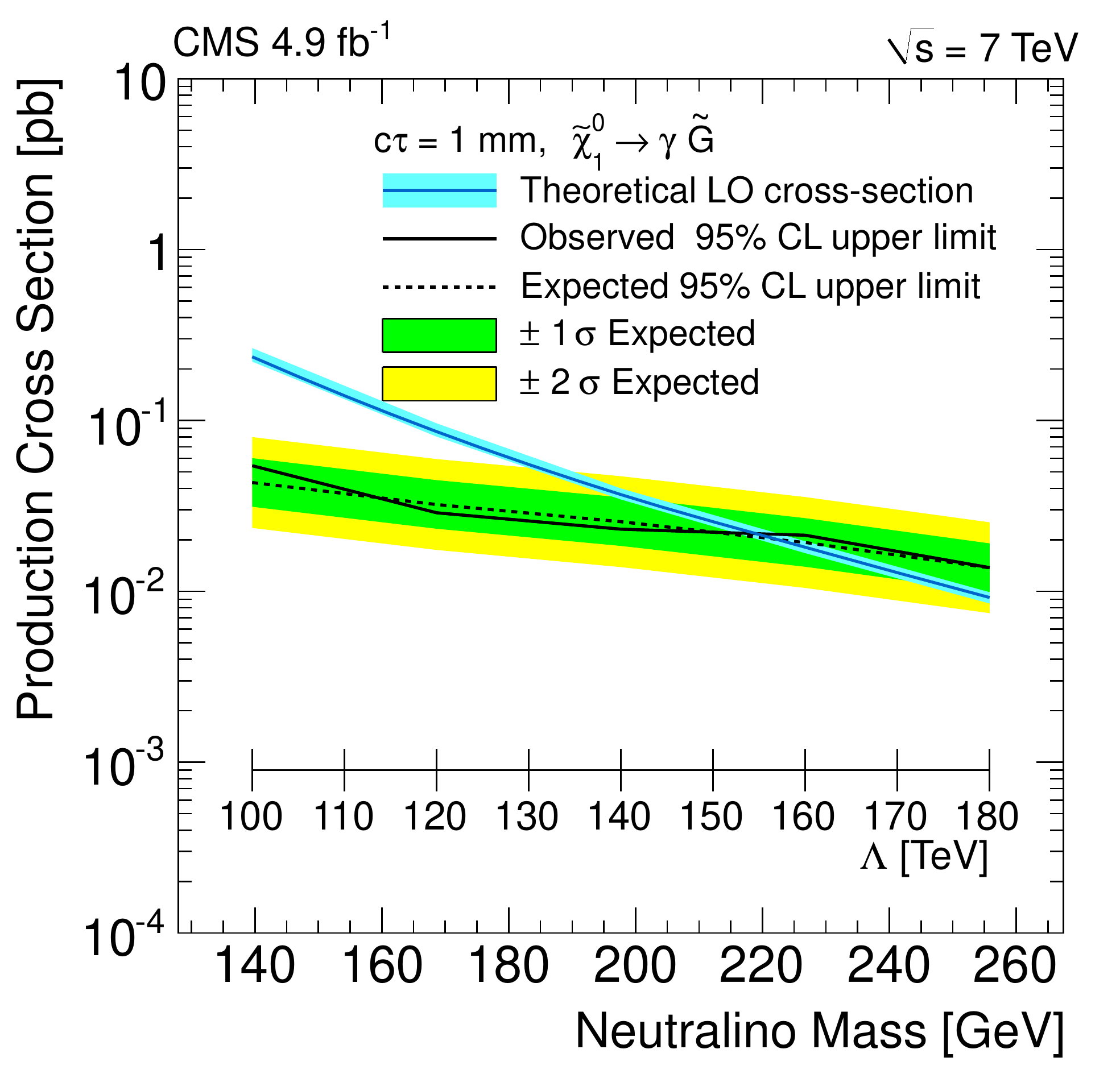}
    \includegraphics[width=\cmsFigWidth]{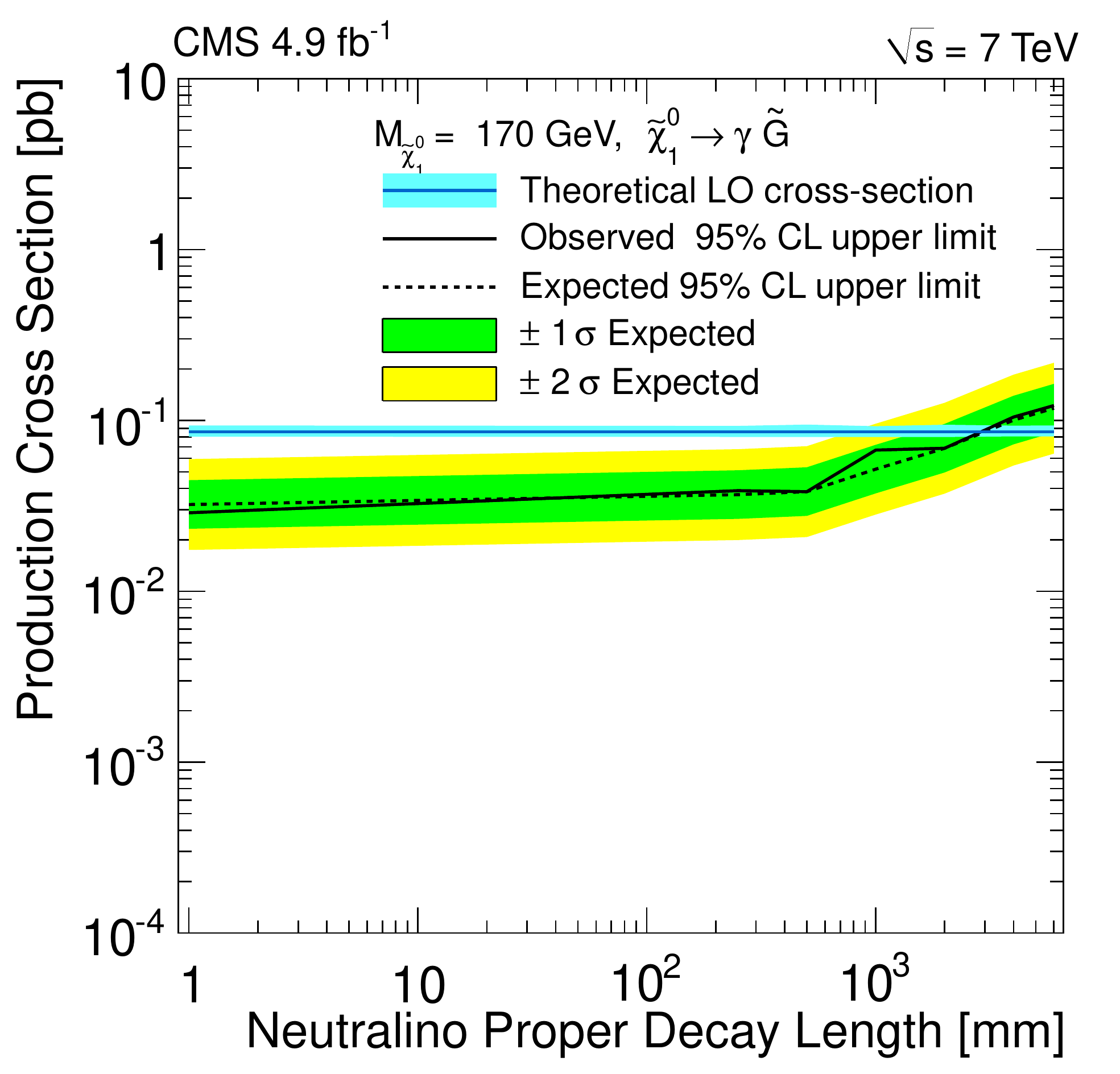}
        \caption{Upper limits at the 95\% CL on the cross section as a function of the
                $\PSGczDo$ mass for $c\tau=1$\unit{mm} (\cmsLeft),
                and for the $\PSGczDo$ proper decay length for $M_{\PSGczDo}=170$\GeV (\cmsRight)
                 in the SPS8 model of GMSB supersymmetry.}
    \label{fig:upperlimit1}
  \end{center}
\end{figure}

\begin{figure}[tbh]
  \begin{center}
    \includegraphics[width=\cmsFigWidth]{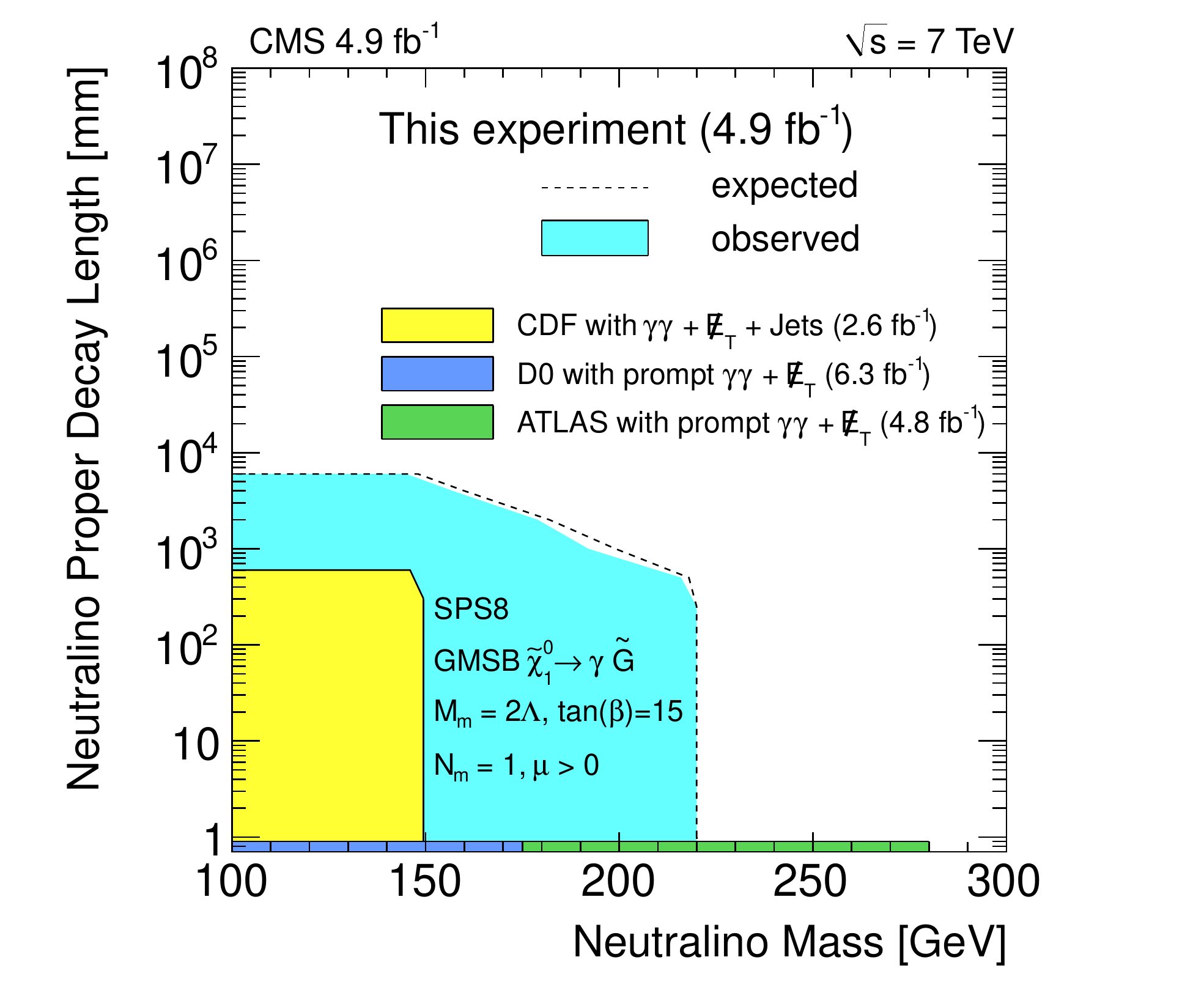}
        \caption{The observed exclusion region for the mass and proper decay length of the long-lived \PSGczDo  in the SPS8 model of GMSB supersymmetry.}
    \label{fig:exclusion}
  \end{center}
\end{figure}

The observed event yield in data is consistent with the SM background
prediction, and upper limits are obtained on the production cross section of a long-lived neutralino
in the context of the GMSB model, assuming $\mathcal{B}(\PSGczDo\rightarrow\gamma\PXXSG)=100\%$.
Exclusion limits are computed with a modified frequentist $\text{CL}_\text{s}$
method~\cite{Higgscombi11,Junk:1999kv,Read:2002hq},
using the asymptotic approximation for the test statistic as described in Ref.~\cite{Cowan:2010js}.
The background normalization and the corresponding uncertainty are taken from the fit
to the data.
The uncertainties in the shapes are taken into account by vertical interpolation of the templates.
The shapes are interpolated  quadratically for shifts below one standard deviation and linearly beyond.
Log-normal multiplicative corrections are used for the normalization, the signal acceptance, and the
integrated luminosity.
Fig.~\ref{fig:upperlimit1} shows the observed and expected 95\% confidence level (CL) upper limits
on the cross section for
GMSB production in terms of $\PSGczDo$ mass (\cmsLeft), and proper decay length (\cmsRight).
The signal cross section is computed at leading order precision and the theoretical uncertainty is evaluated by
using the PDF4LHC recommendation for the PDF uncertainty.
The one-dimensional limits are combined to provide exclusion limits in the mass and proper decay length plane of the long-lived $\PSGczDo$
in Fig.~\ref{fig:exclusion}.

\section{Summary}
The CMS experiment has performed a search for long-lived particles produced in association with jets
using LHC proton-proton collision data at a center-of-mass energy of 7\TeV corresponding to an integrated
luminosity of \lumin. A GMSB scenario with a long-lived neutralino decaying to a photon and a gravitino is
used as the reference.
The missing transverse energy and the timing information from the ECAL
are used to search for an excess of events over the expected SM background prediction.
A fit to the two-dimensional distribution in these variables yields no
significant excess of events beyond the SM contributions, and upper
limits at 95\% CL are obtained on the GMSB
production cross section in the SPS8 model of GMSB supersymmetry.
In this scheme, we obtain an exclusion region as a function of both the neutralino mass and its proper
decay length, assuming $\mathcal{B}(\PSGczDo\rightarrow\gamma\PXXSG)=100\%$.
The mass of the lightest neutralino is then
restricted to values $m(\PSGczDo)>220$\GeV (for neutralino proper decay length $c\tau<500$\unit{mm}) at 95\% CL,
and the neutralino decay length $c\tau$ must be greater than $6000$\unit{mm} (for $m(\PSGczDo)<150$\GeV).
These limits are the most stringent for long-lived neutralinos.

\section*{Acknowledgements}
We congratulate our colleagues in the CERN accelerator departments for the excellent performance of the LHC and thank the technical and administrative staffs at CERN and at other CMS institutes for their contributions to the success of the CMS effort. In addition, we gratefully acknowledge the computing centres and personnel of the Worldwide LHC Computing Grid for delivering so effectively the computing infrastructure essential to our analyses. Finally, we acknowledge the enduring support for the construction and operation of the LHC and the CMS detector provided by the following funding agencies: BMWF and FWF (Austria); FNRS and FWO (Belgium); CNPq, CAPES, FAPERJ, and FAPESP (Brazil); MEYS (Bulgaria); CERN; CAS, MoST, and NSFC (China); COLCIENCIAS (Colombia); MSES (Croatia); RPF (Cyprus); MoER, SF0690030s09 and ERDF (Estonia); Academy of Finland, MEC, and HIP (Finland); CEA and CNRS/IN2P3 (France); BMBF, DFG, and HGF (Germany); GSRT (Greece); OTKA and NKTH (Hungary); DAE and DST (India); IPM (Iran); SFI (Ireland); INFN (Italy); NRF and WCU (Republic of Korea); LAS (Lithuania); CINVESTAV, CONACYT, SEP, and UASLP-FAI (Mexico); MSI (New Zealand); PAEC (Pakistan); MSHE and NSC (Poland); FCT (Portugal); JINR (Armenia, Belarus, Georgia, Ukraine, Uzbekistan); MON, RosAtom, RAS and RFBR (Russia); MSTD (Serbia); SEIDI and CPAN (Spain); Swiss Funding Agencies (Switzerland); NSC (Taipei); ThEPCenter, IPST and NSTDA (Thailand); TUBITAK and TAEK (Turkey); NASU (Ukraine); STFC (United Kingdom); DOE and NSF (USA).

\bibliography{auto_generated}   

\cleardoublepage \appendix\section{The CMS Collaboration \label{app:collab}}\begin{sloppypar}\hyphenpenalty=5000\widowpenalty=500\clubpenalty=5000\textbf{Yerevan Physics Institute,  Yerevan,  Armenia}\\*[0pt]
S.~Chatrchyan, V.~Khachatryan, A.M.~Sirunyan, A.~Tumasyan
\vskip\cmsinstskip
\textbf{Institut f\"{u}r Hochenergiephysik der OeAW,  Wien,  Austria}\\*[0pt]
W.~Adam, E.~Aguilo, T.~Bergauer, M.~Dragicevic, J.~Er\"{o}, C.~Fabjan\cmsAuthorMark{1}, M.~Friedl, R.~Fr\"{u}hwirth\cmsAuthorMark{1}, V.M.~Ghete, J.~Hammer, N.~H\"{o}rmann, J.~Hrubec, M.~Jeitler\cmsAuthorMark{1}, W.~Kiesenhofer, V.~Kn\"{u}nz, M.~Krammer\cmsAuthorMark{1}, I.~Kr\"{a}tschmer, D.~Liko, I.~Mikulec, M.~Pernicka$^{\textrm{\dag}}$, B.~Rahbaran, C.~Rohringer, H.~Rohringer, R.~Sch\"{o}fbeck, J.~Strauss, A.~Taurok, W.~Waltenberger, G.~Walzel, E.~Widl, C.-E.~Wulz\cmsAuthorMark{1}
% ...
\vskip\cmsinstskip
\dag:~Deceased\\
1:~~Also at Vienna University of Technology, Vienna, Austria\\
% ...

\end{sloppypar}
\end{document}